\documentclass[%
 aip,
 jap,%
 amsmath,amssymb,
 reprint,%
]{revtex4-1}

\usepackage{graphicx}
\usepackage{dcolumn}
\usepackage{bm}

\begin{document}

\title{Radiative Heat Transfer in Anisotropic Many-Body Systems: Tuning and Enhancement}
\author{Moladad Nikbakht}
\email{mnik@znu.ac.ir}
\affiliation{Department of Physics, Faculty of Sciences, University of Zanjan, Zanjan 45371-38791, Iran.}
\date{\today}

\begin{abstract}
A general formalism for calculating the  Radiative Heat Transfer in many body systems with anisotropic component is presented. Our scheme extends the theory of radiative heat transfer in isotropic many body systems to anisotropic cases. In addition, the radiative heating of the particles by the thermal bath is taken into account in our formula. It is shown that the radiative heat exchange (HE) between anisotropic particles and their radiative cooling/heating (RCH) could be enhanced several order of magnitude than that of isotropic particles. Furthermore, we demonstrate that both the HE and RCH can be tuned dramatically by particles relative orientation in many body systems.  
\end{abstract}
\maketitle
Nanoscale novel devices have raised the demand for thermal characterization that is critical for device performance and durability. Recent years have seen a rapid growth of interest by scientists in the field of Radiative Heat Transfer (RHT) in nanoscale systems. Theoretical predictions confirmed by experiments have shown that the RHT increases strongly when the distance separating the particles become smaller than the thermal wavelength~\cite{narayanaswamy,kittel,Volokitin,Nara,rousseau,Ottens}. In spite of the great successes of the well-known blackbody radiation formula for RHT at thermal equilibrium, it is only an approximation when we are dealing with non-equilibrium systems. Moreover, such an approximation is reasonable for far apart particles and strongly deviate from the true behavior when they are structured on nanometre scales at which the contribution of evanescent modes are dominate. The Fluctuation-Dissipation Theorem is of relevance for the understanding of fluctuating fields near nanoscale particles and RHT at nanoscale distances~\cite{Manjavacas}. Over the past few years, this approach has been successfully applied to a variety of configurations to calculate the heat exchange (HE)~\cite{messina,huth,ben3,ben1,phan}. The capability and applicability of all these promising studies can be largely enhanced if some degree of tunability is added to break the spatial and/or orientational symmetry in the system~\cite{ben1,biehs1}.   Recently, several studies have been performed on the possibility of controlling the RHT in nanoscale using these factors~\cite{zwol1,zwol2,ben2,basu,Yannopapas,Yannopapas1}. This property can be used for the design of innovative structures to manage the RHT for practical application such as thermophotovoltaic device~\cite{dimatteo,narayanaswamy1,lenert} and thermal sensing~\cite{wilde,huth1,biehs2,jones}. 

In this paper, we present a general formalism for the RHT in a many-body system with anisotropic particles. The system consists of a finite number of anisotropic particles positioned at ${\bm{r}}_i$ inside a thermal bath which is maintained at temperature $T_b$. We wish to obtain a general expression for the HE between these particles (as well as the heat flow between the particles and the bath) when they are maintained at different temperatures ${T_{i}}$. The problem becomes essentially one of determining the amount of power which is dissipated inside each particle by the fluctuating fields, results in RCH of the particles. In general, the RCH of each particle contains contributions from fluctuating fields generated by itself, other particles and the thermal bath. A unique feature of the letter is its emphasis on the tunability of the HE between anisotropic particles by taking into consideration particle-bath heat transfer and many body collective effects. Although we explore the results for the special case of an ellipsoidal nanoparticles, this formalism opens up a powerful route to explore the heat flow in many body systems with arbitrary shaped (and/or polarizability tensor) objects, which may ultimately have a profound impact upon nanoscale devices.
 
The local electric field for an arbitrarily oriented particle, located at ${\bm{r}}_i$ in the system, is determined by~\cite{messina1}
\begin{equation}
{\bf E}_i={\bf E}_i^b+(k^2/\epsilon _0)\sum _{j=1}^N{\hat{\bf G}}_{ij}{\bf P}_j~,
\label{eq1}
\end{equation}
where ${\bf E}_i^b$ is the fluctuating field of the bath, ${\hat{\bf G}}_{i\neq j}$ corresponds to the Green's dyadic in free space, ${\hat{\bf G}}_{ii}={\hat{\bf G}}_{0}=i\frac{k}{6\pi}{\hat{\bf 1}}$, $k=\omega / c$, and particles are taken to be non-polar and non-magnetic. This equation takes an elegant form when written in matrix form,
\begin{equation}
\begin{bmatrix} {\bf E}_1\\ \vdots \\{\bf E}_N
\end{bmatrix}=
\begin{bmatrix} {\bf E}_1^b\\ \vdots \\{\bf E}_N^b
\end{bmatrix}+(k^2/\epsilon_0)
\begin{bmatrix}
{\hat{\bf G}}_{0} & {\hat{\bf G}}_{12} & \cdots & {\hat{\bf G}}_{1N}\\ {\hat{\bf G}}_{21} & {\hat{\bf G}}_{0}& \cdots & {\hat{\bf G}}_{2N}\\ \vdots& \vdots & \ddots & \vdots\\{\hat{\bf G}}_{N1} & {\hat{\bf G}}_{N2}& \cdots &{\hat{\bf G}}_{0}
\end{bmatrix}
\begin{bmatrix} {\bf P}_1\\\vdots \\{\bf P}_N
\end{bmatrix},
\label{eq2}
\end{equation}
which allows for a compact notation
\begin{equation}
{\vec{\bf E}}={\vec{\bf E}^b}+(k^2/\epsilon _0){\hat {\bf {\mathbb G}}}{\vec{\bf P}}.
\label{eq3}
\end{equation}
Since we are dealing with particles (fluctuating dipoles) in a thermal bath, the total electric dipole moment of each particle due to radiative interaction with other particles and the bath can be written as
\begin{equation}
{\bf P}_i={\bf P}_i^{\it ind}+{\bf P}_i^{\it fluc}.
\label{eq4}
\end{equation}
Here, the induced dipole moment ${\bf P}_i^{\it ind}$ of the {\it i}th particle (with polarizability tensor ${\hat{\bm \alpha}}_i$) is connected to the local field  through the relation
\begin{equation}
{\bf P}_i^{\it ind}=\epsilon_0{\hat{\bf \alpha}}_i{\bf E}_i^b+k^2{\hat{\bf \alpha}}_i\sum_{j\neq i}^N{\hat{\bf G}}_{ij}{\bf P}_j.
\label{eq5}
\end{equation}
Using Eqs.~(\ref{eq4}) and (\ref{eq5}), the equation for ${\vec{\bf P}}$ in more compact form is
\begin{equation}
\vec{\bf P}=\vec{\bf P}^{\it fluc}+\epsilon_0{\hat{\bf {\mathbb \alpha}}}{\vec{\bf E}}^b+k^2{\hat{\bf {\mathbb \alpha}}}{\hat {\bf {\mathbb W}}}\vec{\bf P},
\label{eq6}
\end{equation}
In analogy with ${\hat {\mathbb G}}$, ${\hat {\mathbb W}}$ is $3N\times 3N$ block matrix with ${\hat{\bf 0}}$'s along the diagonal and ${\hat{\bf {\mathbb \alpha}}}$ is a block diagonal matrix with ${\hat{\bm \alpha}}_i$ at the $\it i$th block on the diagonal line. Finally, solving for $\vec{\bf P}$ from Eq.~(\ref{eq6}) and then inserting the resultant expression in Eq.~(\ref{eq3}), yields
\begin{subequations}
\begin{eqnarray}
&\vec{\bf P}&={\hat{\mathbb A}}\vec{\bf P}^{\it fluc}+{\hat{\mathbb B}}\vec{\bf E}^b,
\\
&\vec{\bf E}&={\hat{\mathbb C}}\vec{\bf P}^{\it fluc}+{\hat{\mathbb D}}\vec{\bf E}^b,
\end{eqnarray}
\label{eq7}
\end{subequations}
where
\begin{eqnarray}\nonumber
{\hat{\mathbb A}}&=&({\hat{\mathbb I}}-k^2{\hat{\bf {\mathbb \alpha}}} {\hat{\mathbb W}})^{-1}~,~
{\hat{\mathbb B}}=\epsilon_0{\hat{\mathbb A}} {\hat{\bf {\mathbb \alpha}}}, \\
{\hat{\mathbb C}}&=&(k^2/\epsilon_0){\hat{\mathbb G}}{\hat{\mathbb A}}~~~,~~~
{\hat{\mathbb D}}={\hat{\mathbb I}}+k^2{\hat{\mathbb G}}{\hat{\mathbb A}}{\hat{\bf {\mathbb \alpha}}}.
\label{eq8}
\end{eqnarray}
and ${\hat{\mathbb I}}$ is a $3N\times 3N$ identity operator.
The dipole moment ${\bf P}_i$ of the $\it i$th particle interact with the local fields ${\bf E}_i$ such that the total power dissipated in it is given by
\begin{eqnarray}
\mathcal{P}_i=\langle {\bf E}_i^*(t)\cdot{\dot{\bf P}}_i(t)\rangle=2\int_0^\infty\omega\frac{d\omega}{4\pi^2}{\tt Im}{\big[}\langle {\bf E}_i^*(\omega)\cdot{{\bf P}}_i(\omega)\rangle{\big]}.~~~~
\label{eq9}
\end{eqnarray}
According to fluctuation electrodynamics~\cite{Manjavacas}
\begin{subequations}
\begin{eqnarray}
&&\langle {P}_{j',\beta'}^{\it * fluc}\cdot {P}_{j,\beta}^{\it fluc}\rangle=2\pi\hbar\epsilon_0\delta_{jj'}[\frac{1}{2}+n(\omega,T_j)]{\tt Im}({\hat {\bm \chi}}_{j,\beta\beta'}),~~~~~~~~\\
&&\langle {E}_{j',\beta'}^{* b}\cdot {E}_{j,\beta}^{b}\rangle=2\pi\hbar(\frac{k^2}{\epsilon_0})[\frac{1}{2}+n(\omega,T_b)]{\tt Im}({\hat{\bf G}}_{jj',\beta\beta'}).~~~~~~~~
\end{eqnarray}
\label{eq10}
\end{subequations}
where $n(\omega,T)=[\exp(\frac{\hbar \omega}{ k_B T})-1]^{-1}$ is the Bose-Einstein energy distribution function of a quantum oscillator at temperature T and we have introduced ${\hat{\bm \chi}}_j={\hat{\bm \alpha}}_j+k^2{\hat{\bm \alpha}}_j{\hat{\bf G}}_{0}^\dag{\hat{\bm \alpha}}_j^\dag$.
To obtain a convenient expression for the net power dissipated in the {\it i}th particle which results in it's RCH, with the aim of classifying all the possible heat flow that may occur, we can rewrite Eq.~(\ref{eq9}) using Eqs.~(\ref {eq7}) and (\ref{eq10}) in the form
\begin{equation}
\mathcal{P}_i={\mathcal F}_i+\sum_{j \neq i}{\mathcal F}_{i,j}+\sum_{jj'}{\mathcal F}_{i,jj'}^{b}
\label{eq11}
\end{equation}
with
\begin{subequations}
\begin{eqnarray}
&&{\mathcal F}_i={\tt Im}{\int}_0^\infty\frac{d\omega}{\pi}\epsilon_0{\mathbb Tr}[{\hat{\bf A}}_{ii}{\tt Im}({\hat {\bm \chi}}_{i}){\hat{\bf C}}_{ii}^\dag]\Theta(\omega,T_i),~~\\
&&{\mathcal F}_{i,j}={\tt Im}\int_0^\infty\frac{d\omega}{\pi}\epsilon_0{\mathbb Tr}[{\hat{\bf A}}_{ij}{\tt Im}({\hat {\bm \chi}}_{j}){\hat{\bf C}}_{ij}^\dag]\Theta(\omega,T_j),~~\\
&&{\mathcal F}_{i,jj'}^{b}={\tt Im}\int_0^\infty\frac{d\omega}{\pi}(\frac{k^2}{\epsilon_0}){\mathbb Tr}[{\hat{\bf B}}_{ij}{\tt Im}({\hat {\bm G}}_{jj'}){\hat{\bf D}}_{ij'}^\dag]\Theta(\omega,T_b),~~~~~~~
\end{eqnarray}
\label{eq12}
\end{subequations}
and $\Theta(\omega,T)=\hbar\omega[1+2n(\omega,T)]/2$.
${\mathcal F}_{i}$ is the radiative cooling of a particle due to it's radiation in presence of other particles in the system, ${\mathcal F}_{i,j\neq i}$ is radiative heating of the {\it i}th particles due to the radiation of the {\it j}th one, and finally ${\mathcal F}_{i,jj'}^{b}$ stands for the radiative heating of the {\it i}th particles by the thermal bath. Since any radiant energy can be reflected back and forth between the particles several times, all terms in Eq.~(\ref{eq11}) depends on these characteristics including geometrical arrangement, orientations, and shapes. The calculations in Eqs.~(\ref{eq12}) take into consideration these multiple scatterings which is accounted for by the interaction matrixes and polarizability tensors. Moreover, since the absolute values of temperatures are presented in the RCH of each particle, it is expected that the thermal evolution of the particles depends on both the temperatures and temperature differences in the system~\cite{tschikin}.

At this point, we may introduce the HE between two distinct particles ({\it i}th and {\it j}th) in the presence of several scatterers in a system as
\begin{equation}
{\mathcal H}_{ij}=|{\mathcal F}_{i,j}-{\mathcal F}_{j,i}|.
\label{eq13}
\end{equation}
It can be shown that, regardless of temperatures, the net heat flux is always from hotter to colder particle and it vanishes if both particles are at the same temperature, {\it i.e}, $T_i=T_j$. However, even in such a case, these particles might undergo different thermal evolution before thermalizing by bath, because the dynamical behavior is determined by $\mathcal P_i$ rather than $\mathcal H_{ij}$.

For isotropic spherical particles  ${\hat{\bm \alpha}}_i=\alpha_i{\hat{ \bf1}}$ is a scalar quantity, while in most cases like small isotropic nonspherical particles (or different extensions, including anisotropic spherical particles or anisotropic nonspherical particles), it is not a scalar quantity. For special case of small ellipsoidal nanoparticles, the polarizability tensors are diagonal in the principal-axis system of each particle, with diagonal elements given by
\begin{equation}
{\hat\alpha}_{i,\beta\beta}^*=V_i\frac{\varepsilon (\omega)-1}{1+L_{\beta}[\varepsilon(\omega)-1]}, (\beta=x,y,z).
\label{eq14}
\end{equation}
$V_i=\frac{4}{3}\pi a_i b_i c_i$ is the volume of the i{\it th} ellipsoidal particle, $L_{\beta}$ is the depolarization factor and $a_i,~b_i,~c_i$ are the semi-axis of ellipsoid.
Moreover, the radiative correction to the polarizability tensor is~\cite{albaladejo}
\begin{equation}
{\hat{\bm \alpha}}_i={\hat{\bm \alpha}}_i^*\bigg\{{\hat{\bf 1}-i\frac{k^3}{6\pi}{\hat{\bm\alpha}}_i^*}\bigg\}^{-1},\label{eq15}
\end{equation}
As it is clear, elements of the polarizability tensor strongly depend on nanoparticle's characteristics, including shape, size, and inclusion composition. The calculated RCH in Eq.~(\ref {eq11}) or HE in Eq.~(\ref{eq13}) make no reference to any basis and in general are basis independent. So, the polarizability tensor of each particle can be calculated in its principal-axis system and then transformed to an arbitrary reference basis. Under such transformation we would expect that heat flows in the system depend on the particles orientation. Furthermore, in the study below we used SiC as a typical material and the corresponding dielectric function is taken from reference~\cite {handbook}.
\begin{figure}[t]
\includegraphics[scale=1]{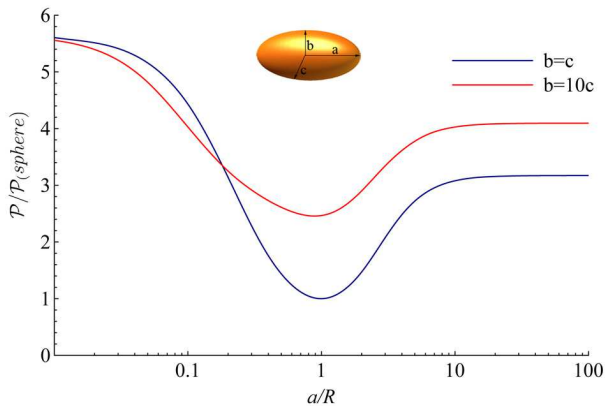}
\caption{(Color online) Radiative cooling/heating of a spheroidal nanoparticle of $a\times b\times c$ in a thermal bath. The particle temperature is $T=350$K while for the thermal bath $T_b=300$K. The particle's volume is kept constant at that of sphere with radius $R$ and the aspect ratio $a/R$ is varied.}
\label{fig1}
\end{figure}

As a preliminary step we have considered the RCH of a single ellipsoidal particle with $T=350$K inside a thermal bath which is maintained at temperature $T_b=300$K. Here, as expected, due to the rotational invariance of Eq.~(\ref {eq11}), the RCH would not depend on particle's orientation. The dependence of the RCH on topological shape of the particle is shown in Fig.~(\ref {fig1}). The volume of the particle is kept constant at that of a sphere with radius $R$ and the ratio $a/R$ is varied. The calculated RCH is normalized to that of spherical particle with the same temperature.

In the case of spheroidal particle, {\it i.e.}, of a rotational ellipsoid with two equal semi-axes, $b=c$, the RCH enhanced when an anisotropy increased, such as growth of nanodisk/nanorod, keeping the volume constant. This is clear, since as the anisotropy increases ({\it i.e.}, from spherical to spheroidal) the threefold degenerate eigen modes with the same contribution in the radiation cooling splits into three eigen modes (one longitude and a twofold degenerate transverse modes) having different eigenfrequencies and contributing with different weights in cooling of the particle. Furthermore, the radiative heating by the bath is proportional to the overlap of the bath radiation spectra and the extinction/absorption cross-section of the particle that would increase or decrease depending on temperatures and the polarizability of the particle. The competition between radiative cooling of the particle and it's heating by bath may be seen more clearly as the anisotropy increases further, that is, going from spheroidal to ellipsoidal shape, ($b\neq c$). When $a/R\lesssim0.2$, the radiative heating of the ellipsoidal particle by the bath increased in comparison to the spheroid particle which results in a small decrease in RCH of the ellipsoidal particle.

Now, consider the energy exchange problem for two anisotropic particles described by $({\bf r}_1,{\hat {\bm \alpha}_1},T_1)$ and $({\bf r}_2,{\hat{\bm \alpha}_2},T_2)$, in a thermal bath at constant temperature $(T_b)$. In this case, each of the nanoparticles exchanges heat with the other and the bath. There would be a net heat flow in such a system as log as $T_1\neq T_2\neq T_b$ which can results in RCH of nanoparticles. We may again apply Eqs.~(\ref{eq11}) and (\ref{eq13}) to calculate the RCH of particles and the heat exchange between them. To evaluate these quantities, the specific characteristics of the system must be known. We shall work out an elementary problem and then present the results for more complicated setups.
\begin{figure}[t]
\includegraphics[scale=.7]{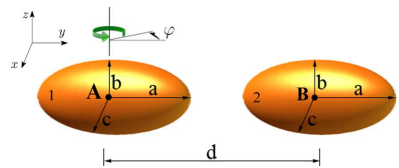}
\caption{\label{fig2} Schematic illustration of a heat transfer in a two body system.}
\end{figure}
Consider, as an example, two identical prolate ellipsoidal nanoparticles (with $a/R=2, b=c$) which separated $d$ apart as shown in Fig.~(\ref{fig2}). The first particle is maintained at $T_1=350$K and the second at $T_2=300$K, while the thermal bath has $T_b=300$K. The calculated heat exchange between the particles versus the orientation of the first particle $\varphi$ are plotted in Fig.~(\ref{fig3}) for various distances. The results are normalized to that of two spheres with same volume and distance.

\begin{figure}[t]
\includegraphics[scale=1]{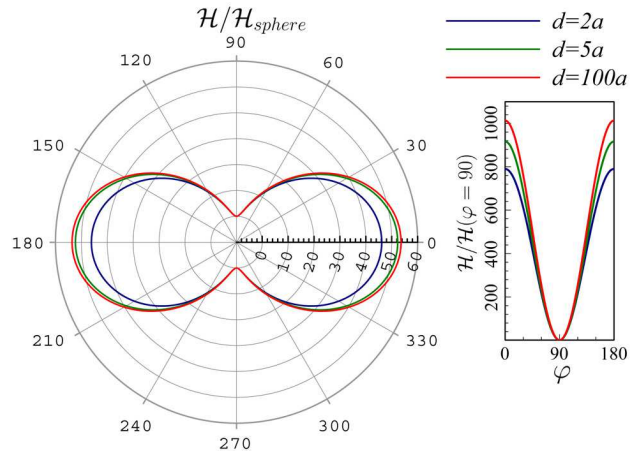}
\caption{\label{fig3} (Color online) Heat exchange between to identical spheroidal particles (separated $d$ apart) with temperatures $T_1=350$ and $T_2=300$, respectively. The temperature of the thermal bath is $T_b=300$ and the calculated heat exchanges are normalized by the value in the system of paire spheres with the same volume. Inset: heat transfer between particles normalized to the the lowest possible energy exchange (in this case $\varphi=90$) for various distances.}
\end{figure}

It is easy to see that the HE strongly enhanced in comparison with the spherical case and maximized for $\varphi=0$. Moreover, as $d$ increased, the contribution of the far-field interaction increased and as a result the HE increased slightly. The beauty of the heat flow in this problem is not only the enhancement but the tunability of the HE between particles. It is interesting that one can amplify the HE between two particles by changing their orientation. This is shown in the infig of Fig.~(\ref{fig3}), where we have represented the HE normalized to it's minimal value ({\it i.e.,} ${\mathcal H}(\varphi=90)$ in this configuration). Evidently, for all distances the curves show amplification in the HE as $\varphi \rightarrow 0$. This amplification corresponds to the change in the polarizability tensor of the first particle as it rotates.

\begin{figure}[t]
\includegraphics[scale=1]{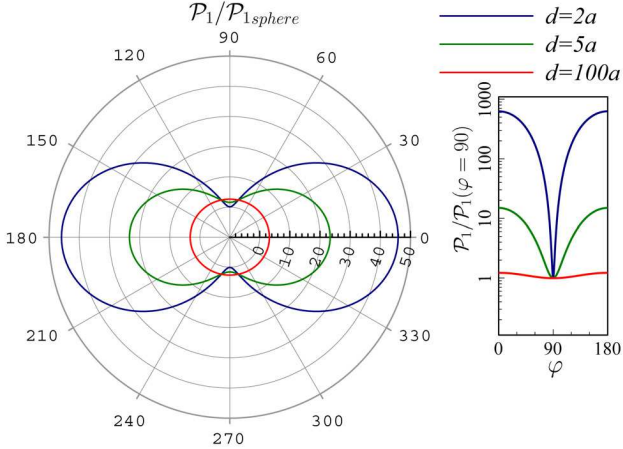}
\caption{\label{fig4} (Color online) Normalized radiative cooling (RCH) of a hotter particle in the system of to identical spheroidal particles with $T_1=350$K, $T_2=300$K , and $T_b=300$K. Inset: The RCH of the first particle normalized to it's minimal value for each distance as a function of orientation $\varphi$. }
\end{figure}

Figure~(\ref{fig4}) shows the orientation dependence of power dissipated in the first particle. Once again all curves are normalized by the value for spheres with volumes equal to the spheroidal volume. Similar to the HE, the RCH is enhanced in comparison with the spherical pair of particles and is maximized in parallel configuration. At small distances, the contribution of the radiative heating by the bath and second particle decreased in comparison to the isotropic particles which results in the increase of the normalized RCH. For large distances, as expected, the RCH losses it's directional dependence and approaches the RCH of a single particle in a bath. The inset of Fig.~(\ref{fig4}) shows the RCH of the first particle normalized to it's minimal value in this configuration. One can see the intensive dependence of the RCH on the orientation of 
the particles as the distances decreases.

\begin{figure}[t]
\includegraphics[scale=1]{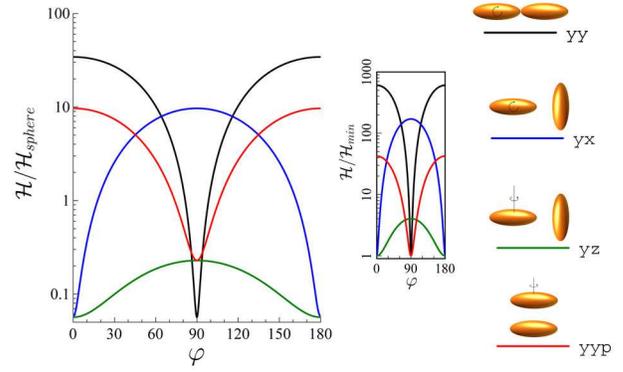}
\caption{\label{fig5} (Color online) Normalized heat exchange between two spheroidal nanoparticles ($a\times b \times b$) as a function of the rotation angle of first particle with parameters $T_1=350$K, $T_2=300$K, $a=4R$, and $d=2a$. Inset: The heat exchange between two spheroidal nanoparticles normalized by the minimal value in each configuration.}
\end{figure}

The calculation of the heat flow in two-body systems may be extended to more complicated geometrical arrangements and particle shapes. For our purposes we give only the results of a few arrangements as shown in Fig.~(\ref{fig5}). Once again, both particles are assumed to be spheroidal with $a=4R$ and separated $d=2a$ apart. In each case, the first particle rotates by angle $\varphi$ while the second one is fixed as shown in Fig.~(\ref{fig5}). The HE is calculated and the results are normalized to that of two spherical particles with same volume, distance and thermal conditions. It can be seen that the extremums of the HE occur when the principal-axes of the particles coincide such as ``end-to-side" ($\varphi_{yy}=90$, $\varphi_{yx}=0$ or $\varphi_{yz}=0$), ``cross-like" ($\varphi_{yz}=90$ or $\varphi_{yyp}=90$), ``side-to-side" ($\varphi_{yx}=90$ or $\varphi_{yyp}=0$) and ``end-to-end" ($\varphi_{yy}=0$) configurations. Among these, the maximum HE enhancement occurs for ``end-to-end" configuration while the ``end-to-side" configuration has the maximum reduction. 
A more detailed description of the tunability of HE between two spheroidal nanoparticles is represented in the inset of Fig.~(\ref{fig5}). In this figure, the HE is normalized to the minimal HE in each configuration. We see that the HE in ``side-to-side" configuration is 1 to 2 orders of magnitude larger than the HE in ``cross-like" configuration.
However, the HE can be increased even further ($\sim 3$ order of magnitude) for ``end-to-end" configuration.

\begin{figure}[t]
\includegraphics[scale=1]{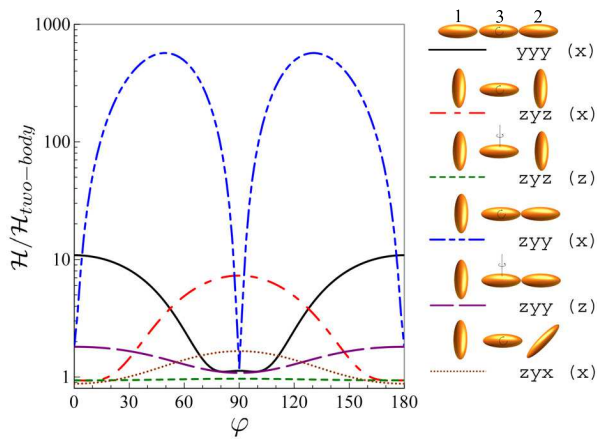}
\caption{\label{fig6} (Color online) Normalized heat exchange between two spheroidal nanoparticles ($a=5b$ , $d_{12}=4a$) maintained at $T_1=350$K (left particle) and $T_2=300$K (right particle) with respect to the orientation of a third one, which is located between the two other particles. The HE is normalized by the HE for two-body system in the same thermal conditions. }
\end{figure}

We have finally considered the radiative heat transfer in three-body systems. For any geometrical configuration, we can use Eq.~(\ref{eq13}) to calculate the interparticle heat exchanges. In such a photon heat transistor, the HE between two particles can be tuned by rotating the third particle. To get an idea of the magnitude of this effect, consider a three aligned spheroidal nanoparticles labeled with indexes 1, 2 and 3 as shown in Fig.~(\ref{fig6}). The third particle is located between the two other particles with equal distances $d_{31}=d_{32}=2a$. Furthermore, we assume that $T_1=350$K, $T_2=300$K and $a=5b$. The position and the orientation of the particles for which the HE is calculated are fixed but the orientation of the third particle is changed by angle $\varphi$ about a given direction. Here, the HE between particles 1 and 2 in a three-body system is calculated and the results are normalized to the HE in the absence of the third particle. It can be seen that the HE between 1 and 2 depends not only on the orientation of the third particle but on their relative orientation.

In the case of ``yyy" configuration, the HE drops an order of magnitude as the third particle rotates about the x (or z) axis to form an ``end-to-side-to-end" configuration. For ``zyz" configuration, the HE is enhanced as the third particle rotates about the x axis and is maximized for ``side-to-side-to-side" configuration while rotation about the z axis plays a negligible role on the HE between 1 and 2. The situation is different in ``zyy" configuration. The orientation dependence is small for rotation of the third particle about the z axis. Interestingly, this dependence is very pronounced for rotation about the x axis. For the Latter case, the HE can dramatically be increased by several orders of magnitude at intermediate values of $\varphi$. Finally, the HE is enhanced in ``zyx" configuration occurs by rotating the third particle by angle $\varphi=90$ about the x (or z) axis. 

In summary, we go deeper into the theory of many-body radiative heat transfer to address particle anisotropy and thermal bath effects. The overall problem of radiative heat transfer is analyzed, including the shape, orientation, and the geometrical arrangement of the particles on their radiative cooling/heating and the total energy that may exchange between them. It is shown that in comparison with isotropic particles, the heat exchange can be enhanced several orders of magnitude for anisotropic particles. Moreover, the radiative cooling/heating of the particles and interparticle heat exchanges can be tuned dramatically by orientation of the particles in the system.

The author thanks P. B. Abdallah and M. Khorrami for helpful discussions.


\begin{thebibliography}{30}%
\makeatletter
\providecommand \@ifxundefined [1]{%
 \@ifx{#1\undefined}
}%
\providecommand \@ifnum [1]{%
 \ifnum #1\expandafter \@firstoftwo
 \else \expandafter \@secondoftwo
 \fi
}%
\providecommand \@ifx [1]{%
 \ifx #1\expandafter \@firstoftwo
 \else \expandafter \@secondoftwo
 \fi
}%
\providecommand \natexlab [1]{#1}%
\providecommand \enquote  [1]{``#1''}%
\providecommand \bibnamefont  [1]{#1}%
\providecommand \bibfnamefont [1]{#1}%
\providecommand \citenamefont [1]{#1}%
\providecommand \href@noop [0]{\@secondoftwo}%
\providecommand \href [0]{\begingroup \@sanitize@url \@href}%
\providecommand \@href[1]{\@@startlink{#1}\@@href}%
\providecommand \@@href[1]{\endgroup#1\@@endlink}%
\providecommand \@sanitize@url [0]{\catcode `\\12\catcode `\$12\catcode
  `\&12\catcode `\#12\catcode `\^12\catcode `\_12\catcode `\%12\relax}%
\providecommand \@@startlink[1]{}%
\providecommand \@@endlink[0]{}%
\providecommand \url  [0]{\begingroup\@sanitize@url \@url }%
\providecommand \@url [1]{\endgroup\@href {#1}{\urlprefix }}%
\providecommand \urlprefix  [0]{URL }%
\providecommand \Eprint [0]{\href }%
\providecommand \doibase [0]{http://dx.doi.org/}%
\providecommand \selectlanguage [0]{\@gobble}%
\providecommand \bibinfo  [0]{\@secondoftwo}%
\providecommand \bibfield  [0]{\@secondoftwo}%
\providecommand \translation [1]{[#1]}%
\providecommand \BibitemOpen [0]{}%
\providecommand \bibitemStop [0]{}%
\providecommand \bibitemNoStop [0]{.\EOS\space}%
\providecommand \EOS [0]{\spacefactor3000\relax}%
\providecommand \BibitemShut  [1]{\csname bibitem#1\endcsname}%
\let\auto@bib@innerbib\@empty
\bibitem [{\citenamefont {Narayanaswamy}, \citenamefont {Shen},\ and\
  \citenamefont {Chen}(2008)}]{narayanaswamy}%
  \BibitemOpen
  \bibfield  {author} {\bibinfo {author} {\bibfnamefont {A.}~\bibnamefont
  {Narayanaswamy}}, \bibinfo {author} {\bibfnamefont {S.}~\bibnamefont {Shen}},
  \ and\ \bibinfo {author} {\bibfnamefont {G.}~\bibnamefont {Chen}},\ }\href
  {\doibase 10.1103/PhysRevB.78.115303} {\bibfield  {journal} {\bibinfo
  {journal} {Phys. Rev. B}\ }\textbf {\bibinfo {volume} {78}},\ \bibinfo
  {pages} {115303} (\bibinfo {year} {2008})}\BibitemShut {NoStop}%
\bibitem [{\citenamefont {Kittel}\ \emph {et~al.}(2005)\citenamefont {Kittel},
  \citenamefont {M\"uller-Hirsch}, \citenamefont {Parisi}, \citenamefont
  {Biehs}, \citenamefont {Reddig},\ and\ \citenamefont {Holthaus}}]{kittel}%
  \BibitemOpen
  \bibfield  {author} {\bibinfo {author} {\bibfnamefont {A.}~\bibnamefont
  {Kittel}}, \bibinfo {author} {\bibfnamefont {W.}~\bibnamefont
  {M\"uller-Hirsch}}, \bibinfo {author} {\bibfnamefont {J.}~\bibnamefont
  {Parisi}}, \bibinfo {author} {\bibfnamefont {S.-A.}\ \bibnamefont {Biehs}},
  \bibinfo {author} {\bibfnamefont {D.}~\bibnamefont {Reddig}}, \ and\ \bibinfo
  {author} {\bibfnamefont {M.}~\bibnamefont {Holthaus}},\ }\href {\doibase
  10.1103/PhysRevLett.95.224301} {\bibfield  {journal} {\bibinfo  {journal}
  {Phys. Rev. Lett.}\ }\textbf {\bibinfo {volume} {95}},\ \bibinfo {pages}
  {224301} (\bibinfo {year} {2005})}\BibitemShut {NoStop}%
\bibitem [{\citenamefont {Volokitin}\ and\ \citenamefont
  {Persson}(2007)}]{Volokitin}%
  \BibitemOpen
  \bibfield  {author} {\bibinfo {author} {\bibfnamefont {A.~I.}\ \bibnamefont
  {Volokitin}}\ and\ \bibinfo {author} {\bibfnamefont {B.~N.~J.}\ \bibnamefont
  {Persson}},\ }\href {\doibase 10.1103/RevModPhys.79.1291} {\bibfield
  {journal} {\bibinfo  {journal} {Rev. Mod. Phys.}\ }\textbf {\bibinfo {volume}
  {79}},\ \bibinfo {pages} {1291} (\bibinfo {year} {2007})}\BibitemShut
  {NoStop}%
\bibitem [{\citenamefont {Narayanaswamy}\ and\ \citenamefont
  {Chen}(2008)}]{Nara}%
  \BibitemOpen
  \bibfield  {author} {\bibinfo {author} {\bibfnamefont {A.}~\bibnamefont
  {Narayanaswamy}}\ and\ \bibinfo {author} {\bibfnamefont {G.}~\bibnamefont
  {Chen}},\ }\href {\doibase 10.1103/PhysRevB.77.075125} {\bibfield  {journal}
  {\bibinfo  {journal} {Phys. Rev. B}\ }\textbf {\bibinfo {volume} {77}},\
  \bibinfo {pages} {075125} (\bibinfo {year} {2008})}\BibitemShut {NoStop}%
\bibitem [{\citenamefont {Rousseau}\ \emph {et~al.}(2009)\citenamefont
  {Rousseau}, \citenamefont {Siria}, \citenamefont {Jourdan}, \citenamefont
  {Volz}, \citenamefont {Comin}, \citenamefont {Chevrier},\ and\ \citenamefont
  {Greffet}}]{rousseau}%
  \BibitemOpen
  \bibfield  {author} {\bibinfo {author} {\bibfnamefont {E.}~\bibnamefont
  {Rousseau}}, \bibinfo {author} {\bibfnamefont {A.}~\bibnamefont {Siria}},
  \bibinfo {author} {\bibfnamefont {G.}~\bibnamefont {Jourdan}}, \bibinfo
  {author} {\bibfnamefont {S.}~\bibnamefont {Volz}}, \bibinfo {author}
  {\bibfnamefont {F.}~\bibnamefont {Comin}}, \bibinfo {author} {\bibfnamefont
  {J.}~\bibnamefont {Chevrier}}, \ and\ \bibinfo {author} {\bibfnamefont
  {J.~J.}\ \bibnamefont {Greffet}},\ }\href {\doibase 10.1038/nphoton.2009.144}
  {\bibfield  {journal} {\bibinfo  {journal} {Nat Photon}\ }\textbf {\bibinfo
  {volume} {3}},\ \bibinfo {pages} {514} (\bibinfo {year} {2009})}\BibitemShut
  {NoStop}%
\bibitem [{\citenamefont {Ottens}\ \emph {et~al.}(2011)\citenamefont {Ottens},
  \citenamefont {Quetschke}, \citenamefont {Wise}, \citenamefont {Alemi},
  \citenamefont {Lundock}, \citenamefont {Mueller}, \citenamefont {Reitze},
  \citenamefont {Tanner},\ and\ \citenamefont {Whiting}}]{Ottens}%
  \BibitemOpen
  \bibfield  {author} {\bibinfo {author} {\bibfnamefont {R.~S.}\ \bibnamefont
  {Ottens}}, \bibinfo {author} {\bibfnamefont {V.}~\bibnamefont {Quetschke}},
  \bibinfo {author} {\bibfnamefont {S.}~\bibnamefont {Wise}}, \bibinfo {author}
  {\bibfnamefont {A.~A.}\ \bibnamefont {Alemi}}, \bibinfo {author}
  {\bibfnamefont {R.}~\bibnamefont {Lundock}}, \bibinfo {author} {\bibfnamefont
  {G.}~\bibnamefont {Mueller}}, \bibinfo {author} {\bibfnamefont {D.~H.}\
  \bibnamefont {Reitze}}, \bibinfo {author} {\bibfnamefont {D.~B.}\
  \bibnamefont {Tanner}}, \ and\ \bibinfo {author} {\bibfnamefont {B.~F.}\
  \bibnamefont {Whiting}},\ }\href {\doibase 10.1103/PhysRevLett.107.014301}
  {\bibfield  {journal} {\bibinfo  {journal} {Phys. Rev. Lett.}\ }\textbf
  {\bibinfo {volume} {107}},\ \bibinfo {pages} {014301} (\bibinfo {year}
  {2011})}\BibitemShut {NoStop}%
\bibitem [{\citenamefont {Manjavacas}\ and\ \citenamefont {Garc\'ia~de
  Abajo}(2012)}]{Manjavacas}%
  \BibitemOpen
  \bibfield  {author} {\bibinfo {author} {\bibfnamefont {A.}~\bibnamefont
  {Manjavacas}}\ and\ \bibinfo {author} {\bibfnamefont {F.~J.}\ \bibnamefont
  {Garc\'ia~de Abajo}},\ }\href {\doibase 10.1103/PhysRevB.86.075466}
  {\bibfield  {journal} {\bibinfo  {journal} {Phys. Rev. B}\ }\textbf {\bibinfo
  {volume} {86}},\ \bibinfo {pages} {075466} (\bibinfo {year}
  {2012})}\BibitemShut {NoStop}%
\bibitem [{\citenamefont {Messina}, \citenamefont {Antezza},\ and\
  \citenamefont {Ben-Abdallah}(2012)}]{messina}%
  \BibitemOpen
  \bibfield  {author} {\bibinfo {author} {\bibfnamefont {R.}~\bibnamefont
  {Messina}}, \bibinfo {author} {\bibfnamefont {M.}~\bibnamefont {Antezza}}, \
  and\ \bibinfo {author} {\bibfnamefont {P.}~\bibnamefont {Ben-Abdallah}},\
  }\href {\doibase 10.1103/PhysRevLett.109.244302} {\bibfield  {journal}
  {\bibinfo  {journal} {Phys. Rev. Lett.}\ }\textbf {\bibinfo {volume} {109}},\
  \bibinfo {pages} {244302} (\bibinfo {year} {2012})}\BibitemShut {NoStop}%
\bibitem [{\citenamefont {{Huth, O.}}\ \emph {et~al.}(2010)\citenamefont
  {{Huth, O.}}, \citenamefont {{Rüting, F.}}, \citenamefont {{Biehs, S.-A.}},\
  and\ \citenamefont {{Holthaus, M.}}}]{huth}%
  \BibitemOpen
  \bibfield  {author} {\bibinfo {author} {\bibnamefont {{Huth, O.}}}, \bibinfo
  {author} {\bibnamefont {{Rüting, F.}}}, \bibinfo {author} {\bibnamefont
  {{Biehs, S.-A.}}}, \ and\ \bibinfo {author} {\bibnamefont {{Holthaus, M.}}},\
  }\href {\doibase 10.1051/epjap/2010027} {\bibfield  {journal} {\bibinfo
  {journal} {Eur. Phys. J. Appl. Phys.}\ }\textbf {\bibinfo {volume} {50}},\
  \bibinfo {pages} {10603} (\bibinfo {year} {2010})}\BibitemShut {NoStop}%
\bibitem [{\citenamefont {Ben-Abdallah}\ \emph {et~al.}(2013)\citenamefont
  {Ben-Abdallah}, \citenamefont {Messina}, \citenamefont {Biehs}, \citenamefont
  {Tschikin}, \citenamefont {Joulain},\ and\ \citenamefont {Henkel}}]{ben3}%
  \BibitemOpen
  \bibfield  {author} {\bibinfo {author} {\bibfnamefont {P.}~\bibnamefont
  {Ben-Abdallah}}, \bibinfo {author} {\bibfnamefont {R.}~\bibnamefont
  {Messina}}, \bibinfo {author} {\bibfnamefont {S.-A.}\ \bibnamefont {Biehs}},
  \bibinfo {author} {\bibfnamefont {M.}~\bibnamefont {Tschikin}}, \bibinfo
  {author} {\bibfnamefont {K.}~\bibnamefont {Joulain}}, \ and\ \bibinfo
  {author} {\bibfnamefont {C.}~\bibnamefont {Henkel}},\ }\href {\doibase
  10.1103/PhysRevLett.111.174301} {\bibfield  {journal} {\bibinfo  {journal}
  {Phys. Rev. Lett.}\ }\textbf {\bibinfo {volume} {111}},\ \bibinfo {pages}
  {174301} (\bibinfo {year} {2013})}\BibitemShut {NoStop}%
\bibitem [{\citenamefont {Ben-Abdallah}, \citenamefont {Biehs},\ and\
  \citenamefont {Joulain}(2011)}]{ben1}%
  \BibitemOpen
  \bibfield  {author} {\bibinfo {author} {\bibfnamefont {P.}~\bibnamefont
  {Ben-Abdallah}}, \bibinfo {author} {\bibfnamefont {S.-A.}\ \bibnamefont
  {Biehs}}, \ and\ \bibinfo {author} {\bibfnamefont {K.}~\bibnamefont
  {Joulain}},\ }\href {\doibase 10.1103/PhysRevLett.107.114301} {\bibfield
  {journal} {\bibinfo  {journal} {Phys. Rev. Lett.}\ }\textbf {\bibinfo
  {volume} {107}},\ \bibinfo {pages} {114301} (\bibinfo {year}
  {2011})}\BibitemShut {NoStop}%
\bibitem [{\citenamefont {Phan}, \citenamefont {Phan},\ and\ \citenamefont
  {Woods}(2013)}]{phan}%
  \BibitemOpen
  \bibfield  {author} {\bibinfo {author} {\bibfnamefont {A.~D.}\ \bibnamefont
  {Phan}}, \bibinfo {author} {\bibfnamefont {T.-L.}\ \bibnamefont {Phan}}, \
  and\ \bibinfo {author} {\bibfnamefont {L.~M.}\ \bibnamefont {Woods}},\ }\href
  {\doibase http://dx.doi.org/10.1063/1.4838875} {\bibfield  {journal}
  {\bibinfo  {journal} {Journal of Applied Physics}\ }\textbf {\bibinfo
  {volume} {114}},\ \bibinfo {eid} {214306} (\bibinfo {year}
  {2013})}\BibitemShut {NoStop}%
\bibitem [{\citenamefont {Biehs}, \citenamefont {Rosa},\ and\ \citenamefont
  {Ben-Abdallah}(2011)}]{biehs1}%
  \BibitemOpen
  \bibfield  {author} {\bibinfo {author} {\bibfnamefont {S.-A.}\ \bibnamefont
  {Biehs}}, \bibinfo {author} {\bibfnamefont {F.~S.~S.}\ \bibnamefont {Rosa}},
  \ and\ \bibinfo {author} {\bibfnamefont {P.}~\bibnamefont {Ben-Abdallah}},\
  }\href {\doibase http://dx.doi.org/10.1063/1.3596707} {\bibfield  {journal}
  {\bibinfo  {journal} {Applied Physics Letters}\ }\textbf {\bibinfo {volume}
  {98}},\ \bibinfo {eid} {243102} (\bibinfo {year} {2011})}\BibitemShut
  {NoStop}%
\bibitem [{\citenamefont {Messina}\ \emph
  {et~al.}(2013{\natexlab{a}})\citenamefont {Messina}, \citenamefont
  {Tschikin}, \citenamefont {Biehs},\ and\ \citenamefont
  {Ben-Abdallah}}]{zwol1}%
  \BibitemOpen
  \bibfield  {author} {\bibinfo {author} {\bibfnamefont {R.}~\bibnamefont
  {Messina}}, \bibinfo {author} {\bibfnamefont {M.}~\bibnamefont {Tschikin}},
  \bibinfo {author} {\bibfnamefont {S.-A.}\ \bibnamefont {Biehs}}, \ and\
  \bibinfo {author} {\bibfnamefont {P.}~\bibnamefont {Ben-Abdallah}},\ }\href
  {\doibase 10.1103/PhysRevB.88.104307} {\bibfield  {journal} {\bibinfo
  {journal} {Phys. Rev. B}\ }\textbf {\bibinfo {volume} {88}},\ \bibinfo
  {pages} {104307} (\bibinfo {year} {2013}{\natexlab{a}})}\BibitemShut
  {NoStop}%
\bibitem [{\citenamefont {van Zwol}\ \emph {et~al.}(2011)\citenamefont {van
  Zwol}, \citenamefont {Joulain}, \citenamefont {Ben~Abdallah}, \citenamefont
  {Greffet},\ and\ \citenamefont {Chevrier}}]{zwol2}%
  \BibitemOpen
  \bibfield  {author} {\bibinfo {author} {\bibfnamefont {P.~J.}\ \bibnamefont
  {van Zwol}}, \bibinfo {author} {\bibfnamefont {K.}~\bibnamefont {Joulain}},
  \bibinfo {author} {\bibfnamefont {P.}~\bibnamefont {Ben~Abdallah}}, \bibinfo
  {author} {\bibfnamefont {J.~J.}\ \bibnamefont {Greffet}}, \ and\ \bibinfo
  {author} {\bibfnamefont {J.}~\bibnamefont {Chevrier}},\ }\href {\doibase
  10.1103/PhysRevB.83.201404} {\bibfield  {journal} {\bibinfo  {journal} {Phys.
  Rev. B}\ }\textbf {\bibinfo {volume} {83}},\ \bibinfo {pages} {201404}
  (\bibinfo {year} {2011})}\BibitemShut {NoStop}%
\bibitem [{\citenamefont {Ben-Abdallah}\ and\ \citenamefont
  {Biehs}(2014)}]{ben2}%
  \BibitemOpen
  \bibfield  {author} {\bibinfo {author} {\bibfnamefont {P.}~\bibnamefont
  {Ben-Abdallah}}\ and\ \bibinfo {author} {\bibfnamefont {S.-A.}\ \bibnamefont
  {Biehs}},\ }\href {\doibase 10.1103/PhysRevLett.112.044301} {\bibfield
  {journal} {\bibinfo  {journal} {Phys. Rev. Lett.}\ }\textbf {\bibinfo
  {volume} {112}},\ \bibinfo {pages} {044301} (\bibinfo {year}
  {2014})}\BibitemShut {NoStop}%
\bibitem [{\citenamefont {Basu}\ and\ \citenamefont {Francoeur}(2011)}]{basu}%
  \BibitemOpen
  \bibfield  {author} {\bibinfo {author} {\bibfnamefont {S.}~\bibnamefont
  {Basu}}\ and\ \bibinfo {author} {\bibfnamefont {M.}~\bibnamefont
  {Francoeur}},\ }\href {\doibase http://dx.doi.org/10.1063/1.3567026}
  {\bibfield  {journal} {\bibinfo  {journal} {Applied Physics Letters}\
  }\textbf {\bibinfo {volume} {98}},\ \bibinfo {eid} {113106} (\bibinfo {year}
  {2011})}\BibitemShut {NoStop}%
\bibitem [{\citenamefont {Yannopapas}\ and\ \citenamefont
  {Vitanov}(2013)}]{Yannopapas}%
  \BibitemOpen
  \bibfield  {author} {\bibinfo {author} {\bibfnamefont {V.}~\bibnamefont
  {Yannopapas}}\ and\ \bibinfo {author} {\bibfnamefont {N.~V.}\ \bibnamefont
  {Vitanov}},\ }\href {\doibase 10.1103/PhysRevLett.110.044302} {\bibfield
  {journal} {\bibinfo  {journal} {Phys. Rev. Lett.}\ }\textbf {\bibinfo
  {volume} {110}},\ \bibinfo {pages} {044302} (\bibinfo {year}
  {2013})}\BibitemShut {NoStop}%
\bibitem [{\citenamefont {Yannopapas}(2013)}]{Yannopapas1}%
  \BibitemOpen
  \bibfield  {author} {\bibinfo {author} {\bibfnamefont {V.}~\bibnamefont
  {Yannopapas}},\ }\href {\doibase 10.1021/jp4033639} {\bibfield  {journal}
  {\bibinfo  {journal} {The Journal of Physical Chemistry C}\ }\textbf
  {\bibinfo {volume} {117}},\ \bibinfo {pages} {14183} (\bibinfo {year}
  {2013})}\BibitemShut {NoStop}%
\bibitem [{\citenamefont {DiMatteo}\ \emph {et~al.}(2001)\citenamefont
  {DiMatteo}, \citenamefont {Greiff}, \citenamefont {Finberg}, \citenamefont
  {Young-Waithe}, \citenamefont {Choy}, \citenamefont {Masaki},\ and\
  \citenamefont {Fonstad}}]{dimatteo}%
  \BibitemOpen
  \bibfield  {author} {\bibinfo {author} {\bibfnamefont {R.~S.}\ \bibnamefont
  {DiMatteo}}, \bibinfo {author} {\bibfnamefont {P.}~\bibnamefont {Greiff}},
  \bibinfo {author} {\bibfnamefont {S.~L.}\ \bibnamefont {Finberg}}, \bibinfo
  {author} {\bibfnamefont {K.~A.}\ \bibnamefont {Young-Waithe}}, \bibinfo
  {author} {\bibfnamefont {H.~K.~H.}\ \bibnamefont {Choy}}, \bibinfo {author}
  {\bibfnamefont {M.~M.}\ \bibnamefont {Masaki}}, \ and\ \bibinfo {author}
  {\bibfnamefont {C.~G.}\ \bibnamefont {Fonstad}},\ }\href {\doibase
  http://dx.doi.org/10.1063/1.1400762} {\bibfield  {journal} {\bibinfo
  {journal} {Applied Physics Letters}\ }\textbf {\bibinfo {volume} {79}},\
  \bibinfo {pages} {1894} (\bibinfo {year} {2001})}\BibitemShut {NoStop}%
\bibitem [{\citenamefont {Narayanaswamy}\ and\ \citenamefont
  {Chen}(2003)}]{narayanaswamy1}%
  \BibitemOpen
  \bibfield  {author} {\bibinfo {author} {\bibfnamefont {A.}~\bibnamefont
  {Narayanaswamy}}\ and\ \bibinfo {author} {\bibfnamefont {G.}~\bibnamefont
  {Chen}},\ }\href {\doibase http://dx.doi.org/10.1063/1.1575936} {\bibfield
  {journal} {\bibinfo  {journal} {Applied Physics Letters}\ }\textbf {\bibinfo
  {volume} {82}},\ \bibinfo {pages} {3544} (\bibinfo {year}
  {2003})}\BibitemShut {NoStop}%
\bibitem [{\citenamefont {Lenert}\ \emph {et~al.}(2014)\citenamefont {Lenert},
  \citenamefont {Bierman}, \citenamefont {Nam}, \citenamefont {Celanovic},
  \citenamefont {Soljacic},\ and\ \citenamefont {Wang}}]{lenert}%
  \BibitemOpen
  \bibfield  {author} {\bibinfo {author} {\bibfnamefont {A.}~\bibnamefont
  {Lenert}}, \bibinfo {author} {\bibfnamefont {D.~M.}\ \bibnamefont {Bierman}},
  \bibinfo {author} {\bibfnamefont {W.~R.}\ \bibnamefont {Nam}, \bibfnamefont
  {Y.~Chan}}, \bibinfo {author} {\bibfnamefont {I.}~\bibnamefont {Celanovic}},
  \bibinfo {author} {\bibfnamefont {M.}~\bibnamefont {Soljacic}}, \ and\
  \bibinfo {author} {\bibfnamefont {E.~N.}\ \bibnamefont {Wang}},\ }\href
  {\doibase 10.1038/nnano.2013.286} {\bibfield  {journal} {\bibinfo  {journal}
  {Nat Nano}\ }\textbf {\bibinfo {volume} {9}},\ \bibinfo {pages} {126}
  (\bibinfo {year} {2014})}\BibitemShut {NoStop}%
\bibitem [{\citenamefont {De~Wilde}\ \emph {et~al.}(2006)\citenamefont
  {De~Wilde}, \citenamefont {Formanek}, \citenamefont {Carminati},
  \citenamefont {Gralak}, \citenamefont {Lemoine}, \citenamefont {Joulain},
  \citenamefont {Mulet}, \citenamefont {Chen},\ and\ \citenamefont
  {Greffet}}]{wilde}%
  \BibitemOpen
  \bibfield  {author} {\bibinfo {author} {\bibfnamefont {Y.}~\bibnamefont
  {De~Wilde}}, \bibinfo {author} {\bibfnamefont {F.}~\bibnamefont {Formanek}},
  \bibinfo {author} {\bibfnamefont {R.}~\bibnamefont {Carminati}}, \bibinfo
  {author} {\bibfnamefont {B.}~\bibnamefont {Gralak}}, \bibinfo {author}
  {\bibfnamefont {P.~A.}\ \bibnamefont {Lemoine}}, \bibinfo {author}
  {\bibfnamefont {K.}~\bibnamefont {Joulain}}, \bibinfo {author} {\bibfnamefont
  {J.~P.}\ \bibnamefont {Mulet}}, \bibinfo {author} {\bibfnamefont
  {Y.}~\bibnamefont {Chen}}, \ and\ \bibinfo {author} {\bibfnamefont {J.~J.}\
  \bibnamefont {Greffet}},\ }\href {\doibase 10.1038/nature05265} {\bibfield
  {journal} {\bibinfo  {journal} {Nature}\ }\textbf {\bibinfo {volume} {444}},\
  \bibinfo {pages} {740} (\bibinfo {year} {2006})}\BibitemShut {NoStop}%
\bibitem [{\citenamefont {Huth}\ \emph {et~al.}(2011)\citenamefont {Huth},
  \citenamefont {Schnell}, \citenamefont {Wittborn}, \citenamefont {Ocelic},\
  and\ \citenamefont {Hillenbrand}}]{huth1}%
  \BibitemOpen
  \bibfield  {author} {\bibinfo {author} {\bibfnamefont {F.}~\bibnamefont
  {Huth}}, \bibinfo {author} {\bibfnamefont {M.}~\bibnamefont {Schnell}},
  \bibinfo {author} {\bibfnamefont {J.}~\bibnamefont {Wittborn}}, \bibinfo
  {author} {\bibfnamefont {N.}~\bibnamefont {Ocelic}}, \ and\ \bibinfo {author}
  {\bibfnamefont {R.}~\bibnamefont {Hillenbrand}},\ }\href {\doibase
  10.1038/nmat3006} {\bibfield  {journal} {\bibinfo  {journal} {Nat Mater}\
  }\textbf {\bibinfo {volume} {10}},\ \bibinfo {pages} {352} (\bibinfo {year}
  {2011})}\BibitemShut {NoStop}%
\bibitem [{\citenamefont {Biehs}\ \emph {et~al.}(2010)\citenamefont {Biehs},
  \citenamefont {Huth}, \citenamefont {Rüting},\ and\ \citenamefont
  {Holthaus}}]{biehs2}%
  \BibitemOpen
  \bibfield  {author} {\bibinfo {author} {\bibfnamefont {S.-A.}\ \bibnamefont
  {Biehs}}, \bibinfo {author} {\bibfnamefont {O.}~\bibnamefont {Huth}},
  \bibinfo {author} {\bibfnamefont {F.}~\bibnamefont {Rüting}}, \ and\
  \bibinfo {author} {\bibfnamefont {M.}~\bibnamefont {Holthaus}},\ }\href
  {\doibase http://dx.doi.org/10.1063/1.3437651} {\bibfield  {journal}
  {\bibinfo  {journal} {Journal of Applied Physics}\ }\textbf {\bibinfo
  {volume} {108}},\ \bibinfo {eid} {014312} (\bibinfo {year}
  {2010})}\BibitemShut {NoStop}%
\bibitem [{\citenamefont {Jones}\ and\ \citenamefont {Raschke}(2012)}]{jones}%
  \BibitemOpen
  \bibfield  {author} {\bibinfo {author} {\bibfnamefont {A.~C.}\ \bibnamefont
  {Jones}}\ and\ \bibinfo {author} {\bibfnamefont {M.~B.}\ \bibnamefont
  {Raschke}},\ }\href {\doibase 10.1021/nl204201g} {\bibfield  {journal}
  {\bibinfo  {journal} {Nano Letters}\ }\textbf {\bibinfo {volume} {12}},\
  \bibinfo {pages} {1475} (\bibinfo {year} {2012})}\BibitemShut {NoStop}%
\bibitem [{\citenamefont {Messina}\ \emph
  {et~al.}(2013{\natexlab{b}})\citenamefont {Messina}, \citenamefont
  {Tschikin}, \citenamefont {Biehs},\ and\ \citenamefont
  {Ben-Abdallah}}]{messina1}%
  \BibitemOpen
  \bibfield  {author} {\bibinfo {author} {\bibfnamefont {R.}~\bibnamefont
  {Messina}}, \bibinfo {author} {\bibfnamefont {M.}~\bibnamefont {Tschikin}},
  \bibinfo {author} {\bibfnamefont {S.-A.}\ \bibnamefont {Biehs}}, \ and\
  \bibinfo {author} {\bibfnamefont {P.}~\bibnamefont {Ben-Abdallah}},\ }\href
  {\doibase 10.1103/PhysRevB.88.104307} {\bibfield  {journal} {\bibinfo
  {journal} {Phys. Rev. B}\ }\textbf {\bibinfo {volume} {88}},\ \bibinfo
  {pages} {104307} (\bibinfo {year} {2013}{\natexlab{b}})}\BibitemShut
  {NoStop}%
\bibitem [{\citenamefont {{Tschikin, M.}}\ \emph {et~al.}(2012)\citenamefont
  {{Tschikin, M.}}, \citenamefont {{Biehs, S.-A.}}, \citenamefont {{Rosa,
  F.S.S.}},\ and\ \citenamefont {{Ben-Abdallah, P.}}}]{tschikin}%
  \BibitemOpen
  \bibfield  {author} {\bibinfo {author} {\bibnamefont {{Tschikin, M.}}},
  \bibinfo {author} {\bibnamefont {{Biehs, S.-A.}}}, \bibinfo {author}
  {\bibnamefont {{Rosa, F.S.S.}}}, \ and\ \bibinfo {author} {\bibnamefont
  {{Ben-Abdallah, P.}}},\ }\href {\doibase 10.1140/epjb/e2012-30219-7}
  {\bibfield  {journal} {\bibinfo  {journal} {Eur. Phys. J. B}\ }\textbf
  {\bibinfo {volume} {85}},\ \bibinfo {pages} {233} (\bibinfo {year}
  {2012})}\BibitemShut {NoStop}%
\bibitem [{\citenamefont {Albaladejo}\ \emph {et~al.}(2010)\citenamefont
  {Albaladejo}, \citenamefont {G\'{o}mez-Medina}, \citenamefont
  {Froufe-P\'{e}rez}, \citenamefont {Marinchio}, \citenamefont {Carminati},
  \citenamefont {Torrado}, \citenamefont {Armelles}, \citenamefont
  {Garc\'{i}a-Mart\'{i}n},\ and\ \citenamefont {S\'{a}enz}}]{albaladejo}%
  \BibitemOpen
  \bibfield  {author} {\bibinfo {author} {\bibfnamefont {S.}~\bibnamefont
  {Albaladejo}}, \bibinfo {author} {\bibfnamefont {R.}~\bibnamefont
  {G\'{o}mez-Medina}}, \bibinfo {author} {\bibfnamefont {L.~S.}\ \bibnamefont
  {Froufe-P\'{e}rez}}, \bibinfo {author} {\bibfnamefont {H.}~\bibnamefont
  {Marinchio}}, \bibinfo {author} {\bibfnamefont {R.}~\bibnamefont
  {Carminati}}, \bibinfo {author} {\bibfnamefont {J.~F.}\ \bibnamefont
  {Torrado}}, \bibinfo {author} {\bibfnamefont {G.}~\bibnamefont {Armelles}},
  \bibinfo {author} {\bibfnamefont {A.}~\bibnamefont {Garc\'{i}a-Mart\'{i}n}},
  \ and\ \bibinfo {author} {\bibfnamefont {J.~J.}\ \bibnamefont {S\'{a}enz}},\
  }\href {\doibase 10.1364/OE.18.003556} {\bibfield  {journal} {\bibinfo
  {journal} {Opt. Express}\ }\textbf {\bibinfo {volume} {18}},\ \bibinfo
  {pages} {3556} (\bibinfo {year} {2010})}\BibitemShut {NoStop}%
\bibitem [{\citenamefont {Palik}\ and\ \citenamefont {Ghosh}(1998)}]{handbook}%
  \BibitemOpen
  \bibfield  {author} {\bibinfo {author} {\bibfnamefont {E.}~\bibnamefont
  {Palik}}\ and\ \bibinfo {author} {\bibfnamefont {G.}~\bibnamefont {Ghosh}},\
  }\href@noop {} {\emph {\bibinfo {title} {Handbook of optical constants of
  solids}}}\ (\bibinfo  {publisher} {Academic press},\ \bibinfo {year}
  {1998})\BibitemShut {NoStop}%
\end{thebibliography}
%

\end{document}